# Beam steering at the nanosecond time scale with an atomically thin reflector

Trond I. Andersen[1], Ryan J. Gelly[1], Giovanni Scuri[1], Bo L. Dwyer[1], Dominik S. Wild[1,2], Rivka Bekenstein[1,3], Andrey Sushko[1], Jiho Sung[1,4], You Zhou[1,4,5], Alexander A. Zibrov[1], Xiaoling Liu[1], Andrew Y. Joe[1], Kenji Watanabe[6], Takashi Taniguchi[7], Susanne F. Yelin[1], Philip Kim[1,8], Hongkun Park[1,4,†], Mikhail D. Lukin[1,†]

[1]Department of Physics, Harvard University, Cambridge, MA 02138, USA
[2]Max Planck Institute of Quantum Optics, Hans-Kopfermann-Straße 1, D-85748 Garching, Germany
[3]ITAMP, Harvard-Smithsonian Center for Astrophysics, Cambridge, MA 02138, USA
[4]Department of Chemistry and Chemical Biology, Harvard University, Cambridge, MA 02138, USA
[5]Department of Materials Science and Engineering, University of Maryland, College Park, MD, USA
[6]Research Center for Functional Materials, National Institute for Materials Science, 1-1 Namiki, Tsukuba 305-0044, Japan
[7]International Center for Materials Nanoarchitectonics, National Institute for Materials Science, 1-1 Namiki, Tsukuba 305-0044, Japan
[8]John A. Paulson School of Engineering and Applied Sciences, Harvard University, Cambridge, MA 02138, USA
[†]To whom correspondence should be addressed: lukin@physics.harvard.edu and hongkun_park@harvard.edu

**Techniques to mold the flow of light on subwavelength scales enable fundamentally new optical systems and device applications. The realization of programmable, active optical systems with fast, tunable components is among the outstanding challenges in the field. Here, we experimentally demonstrate a few-pixel beam steering device based on electrostatic gate control of excitons in an atomically thin semiconductor with strong light-matter interactions. By combining the high reflectivity of a $MoSe_2$ monolayer with a graphene split-gate geometry, we shape the wavefront phase profile to achieve continuously tunable beam deflection with a range of 10°, two-dimensional beam steering, and switching times down to 1.6 nanoseconds. Our approach opens the door for a new class of atomically thin optical systems, such as rapidly switchable beam arrays and quantum metasurfaces operating at their fundamental thickness limit.**

**Introduction**

Conventional optical devices, typically made from materials with relatively weak light-matter interactions and a smooth optical response on the wavelength scale, are bulky to accumulate the desired effect on the optical wavefront. Recent advances in flat optics demonstrate that steep gradients in the phase, amplitude or polarization can be used to control light fields on sub-wavelength scales[1-3], enabling novel optical phenomena and applications, including ultrathin lenses[4,5], metasurfaces[6], non-reciprocity[7], and negative refraction[8].

While most demonstrations of flat optical elements so far have involved passive devices, their active counterparts have recently attracted great interest[9-12]. Tuning mechanisms including optically[13] and thermally[14] induced phase transitions, as well as magnetically tuned transparency in magneto-plasmonic crystals[15], have been demonstrated. Micro-electrical mechanical systems (MEMS) technology has also been employed to spatially modulate the optical response[16], but the operation speed of such devices is typically limited to the kHz or few MHz range. A promising avenue to overcome this limitation involves full electrical control, via for instance ionic[17] or electrostatic[18-20] gating. In order to achieve fully programmable devices, a key challenge is to achieve fast, continuous tunability of multiple independent channels.

To address this challenge, we realize a continuously tunable, atomically thin optical device based on phase profile modulation in field-effect transistors composed entirely of two-dimensional van der Waals materials. The optically active element of our system is exfoliated monolayer $MoSe_2$ – an atomically thin semiconductor that hosts tightly bound excitons in the optical (visible) domain. Excitons in TMDs have been widely proposed as an appealing system for quantum optical devices[21], due to their quantum coherence properties[22] and potential for quantum nonlinear effects mediated by confinement-enhanced exciton-exciton interactions[23]. In high-quality exfoliated flakes encapsulated in boron nitride (hBN), these excitons can exhibit very strong light-matter interaction, enabling almost perfect reflection from an atomically thin reflector[24-26]. By employing near-transparent graphene gates, the exciton resonance in TMDs can be electrostatically tuned[27,28]. Due to the two-dimensional nature of TMDs, the exciton resonance is tuned throughout the whole material, circumventing the effects of screening commonly encountered in bulk semiconductors[29]. These features allow for spatially dependent control of the phase and amplitude of the reflected light, which modifies the beam in the far-field due to interference, enabling wide-ranging possibilities for optical beam control (Fig. 1a).

## Results

### Phase modulation through electrostatic gating of excitons

We demonstrate this approach by realizing fast, continuously tunable beam steering with a split-gate geometry (Fig. 1b). Using the dry-transfer technique, we assemble our device (Fig. 1b, inset) from exfoliated flakes of monolayer $MoSe_2$, graphene (top and bottom gates), and hBN (gate dielectric) into a split-gate field effect transistor (SG-FET) structure, in which the bottom gate (BG) only covers part of the device. The gate geometry enables independent electrostatic doping of the two parts of the device, and – because the exciton resonance shifts with doping – a very steep phase gradient near the gate edge. The non-zero width of this step is due to stray electric fields and is comparable to the thickness of the gate dielectric (~50 nm). By focusing light on the gate-edge, the two halves of the reflected wavefront gain a different phase ($\Delta\phi$) and thus constructively interfere in the far-field at an angle to the optical axis (Fig. 1b).

Figure 1c shows the absolute reflectance spectrum, collected away from the gate edge in the intrinsic regime ($T \sim 6$ K). The $MoSe_2$ features a sharp excitonic resonance at $\lambda_{\text{intrinsic}}$ ~757 nm, with a linewidth of 1.2 nm. The asymmetric line-shape is attributed to the interference between the Lorentzian exciton reflection and light reflected from other interfaces in the device[26] (Supplementary Notes I and II). By fitting the asymmetric resonance profile (black; Supplementary Note I), we extract the phase of the reflected light (red). The phase of the exciton reflection itself changes by 180° ($\pi$) across the resonance, but the interference with the background reflections reduces the overall phase range (we note that this is not an inherent limitation; see Supplementary Note II and Supplementary Figures S1-S3).

In order to tune the phase at a given wavelength, we shift the exciton resonance by electrostatically doping the MoSe$_2$ with the top and bottom gates ($V_{TG}$ and $V_{BG}$, respectively). Figures 1d and e show the gate-dependence of the resonance wavelengths, $\lambda_{left}$ and $\lambda_{right}$, on the left and right sides of the gate edge, respectively. The exciton resonance blue-shifts by several linewidths in the p- and n-doped regimes, resulting from bandgap renormalization and repulsive polaron formation[26,28,30]. In the left (dual-gated) side of the device (Fig. 1d), the resonance depends on both $V_{TG}$ and $V_{BG}$; more specifically the weighted sum, $8V_{TG}+V_{BG}$, due to unequal top and bottom hBN thicknesses. In contrast, the exciton resonance on the right (single-gated) side of the device only depends on $V_{TG}$, except for a slight $V_{BG}$-dependence of the onset of the p-doped regime due to contact activation (Fig. 1e). These distinct dependencies of $\lambda_{left}$ and $\lambda_{right}$ on the gate voltages allow for tuning their relative positioning (Fig. 1f) and are key to creating the abrupt phase discontinuity.

Fitting the spectra at all gate voltage combinations in Figs. 1d and e, we find the gate dependence of the phase difference ($\Delta\phi$) between the two sides of the device at $\lambda_0$ =755.6 nm (Fig. 1g), indicating a continuously tunable range of 42°. The wavelength is chosen to be blue-detuned relative to the intrinsic exciton resonance, such that the exciton resonance is swept through $\lambda_0$ upon electrostatic gating. Inspecting Fig. 1g, we identify four regimes that are central to the operation of our system, distinguished by the relative positioning of $\lambda_{left}$ and $\lambda_{right}$: a large positive phase difference is achieved when the resonance on the left side is blue-shifted past $\lambda_0$, while $\lambda_{right}$ is kept red-shifted (Fig. 1f, blue box). Conversely, a large negative phase difference is realized when $\lambda_{right} < \lambda_0 < \lambda_{left}$ (Fig. 1f, green box). Finally, the magnitude of the phase difference is much smaller when the two sides are either both doped (Fig. 1f, yellow box) or both intrinsic (Fig. 1f, red box). As can be seen in Fig. 1f, the doping induced blue-shift is accompanied by a decrease in amplitude, reducing the maximum phase of the combined reflection in the doped regimes.

**One-dimensional beam steering**

Having demonstrated independent phase tunability in the two sides of the device, we next measure the beam steering capabilities by focusing the laser beam ($\lambda_0$ =755.6 nm; numerical aperture, NA=0.75) onto the gate edge (see Supplementary Note III for alignment procedure) and imaging the reflected beam in the Fourier plane (see Supplementary Video S1). The Fourier plane polar coordinates $r_F$ and $\phi_F$ are converted to angular deflection via $\theta = \sin^{-1}(r_F/f)$, where $f$ is the focal length of the objective, and decomposed into $\theta_x = \theta \cdot \cos(\phi_F)$ and $\theta_y = \theta \cdot \sin(\phi_F)$. The undeflected beam is approximately Gaussian with an angular width (standard deviation) of 17°. Figure 2a shows representative Fourier images in the four regimes identified in Figs. 1f and g, after subtracting the reflected intensity without exciton effects ($I_0$), obtained by heavily doping the device ($V_{BG}$=10 V, $V_{TG}$=1.4 V). Defining the center-of-mass deflection of the full (not background-subtracted) reflection as $\bar{\theta}_i = \frac{\Sigma I \theta_i}{\Sigma I}$, we present scatter plots of $\bar{\theta}_x$ and $\bar{\theta}_y$ for the full range of gate voltages in Fig. 2b, including highlighted deflections in the four regimes (colored circles) and Fourier images without background subtraction (insets).

We find that the reflection is deflected in the expected direction, perpendicular to the gate edge. The tunable deflection range is 9.8°, in very good agreement with theoretical predictions based on the phase difference range observed in Fig. 1g. Specifically, for a diffraction limited beam spot, the deflection is predicted to be $\bar{\theta}_\perp = \frac{NA}{0.42 \cdot \sqrt{2\pi^3}} \cdot \Delta\phi = 0.23 \cdot \Delta\phi$, which gives a range of 9.7° (Supplementary Note IV). Moreover, the gate dependence of the deflection (Fig. 2c) closely resembles that of the extracted phase difference (Fig. 1g), consistent with the deflection arising due to the sharp phase discontinuity imparted on the wavefront. We identify distinct steering behavior in the four regimes: when only one side of the device is kept red-shifted relative to $\lambda_0$ (blue and green boxes in Fig. 2a), the beam is deflected towards that side. Intuitively, the light from the two sides interferes constructively when the path length is shorter for the side with the greater phase. In the two other regimes, on the other hand, where the two sides are either both blue-shifted past $\lambda_0$ (yellow) or both red-shifted (red), we observe near-zero deflection, consistent with a much smaller phase difference. We note that very similar behavior, although with a smaller deflection range, was observed for $\lambda_0$ down to 752 nm.

Figure 2d shows the gate dependence of the integrated reflection. The different regimes are separated by regions of strong reflection, since these indicate that one of the resonances crosses through $\lambda_0$. While the four different regimes are easily identified, we emphasize that the phase difference, and thus the beam deflection, is tuned continuously, as shown in Fig. 2b. The continuous tunability is highlighted in Fig. 2e, showing linecuts from Fig. 2c. Although the reflection amplitude is gate dependent (Figs. 2d,f), the reflection variations can be reduced substantially while still keeping a similar deflection range by avoiding the combined resonance (gray linecut), or even designing a near-iso-reflection path in voltage space (teal).

**Two-dimensional beam steering**

In order to achieve more advanced control of the wavefront profile, we utilize the device region where the edge of the bottom gate intersects a border between monolayer and bilayer MoSe$_2$ (Fig. 3a). Since the exciton resonance in bilayers is red-shifted to ~767 nm due to interlayer hybridization effects[31] (Fig. 3b), the bilayer acts as a non-resonant dielectric reflector, thus enabling control of the relative phase between the three regions. Figure 3c shows representative Fourier images of the reflected light in the four different regimes at $\lambda_0 = 754.1$ nm. When both the monolayer resonances are red-shifted relative to $\lambda_0$, the beam is deflected upwards (red), as opposed to the near-zero deflection observed in Fig. 2. Keeping only one of the resonances red-shifted causes the beam to deflect upwards at an angle towards the red-shifted side (green and blue). This is further shown by plotting the full set of center-of-mass deflections, $\bar{\theta}_x$ and $\bar{\theta}_y$ (Fig. 3d); while these were clustered around a line perpendicular to the gate edge in Fig. 2b, we now observe that they span a broader, two-dimensional area. Consequently, the gate dependence of $\bar{\theta}_x$ and $\bar{\theta}_y$ (Figs. 3e,f) – while still resembling that of the phase in Fig. 1g – is now more intricate. Instead of simply being (negatively) proportional to each other, $\bar{\theta}_x$ takes on both positive and negative values when

$\bar{\theta}_y$ is intermediate, and $\bar{\theta}_y$ can be either large or near-zero when $\bar{\theta}_x$ is near zero. To demonstrate the two-dimensional steering capability, we write a desired two-dimensional pattern with the center-of-mass of the reflected beam (Fig. 3g), by successively applying gate voltage combinations corresponding to the appropriate beam deflections.

The two-dimensional deflection behavior is well understood by considering the third reflection source from the bilayer region. When both the monolayer resonances are kept red-detuned relative to $\lambda_0$, the phase is higher than in the bilayer region, thus imparting an upwards phase gradient on the reflected wavefront. Similarly, if only one of the monolayer regions is kept red-shifted, the phase gradient points towards that region. Hence, the three reflection sources enable two-dimensional beam control.

**Switching on the nanosecond scale**

We investigate the temporal response of our system by applying a small oscillating bottom gate voltage, $V_{BG}(t) = V_0 + \Delta V \cdot \sin\left(\frac{2\pi t}{\tau}\right) = 0.7 \text{ V} + 0.45 \text{ V} \cdot \sin\left(\frac{2\pi t}{\tau}\right)$ and a constant top gate voltage $V_{TG} = 0.64$ V, where $\tau$ is the period (corresponding to twice the switching time). Focusing the beam at the gate edge, as in Fig. 2, we first measure the beam deflection using a long period ($\tau = 2$ s) compatible with our camera (Fig. 4a). Figures 4b and c show the change in reflection from $V_{BG} = V_0$ to $V_{BG} = V_0 + \Delta V$ and $V_{BG} = V_0 - \Delta V$, respectively. Next, we measure the optical response at much higher frequencies by collecting the reflected light in the Fourier plane with an avalanche photon detector (APD; Fig. 4d). In order to ensure that we probe beam steering, as opposed to simply changes in reflection amplitude, we separately collect photons from the left ($\theta_x < 0$) and right parts ($\theta_x > 0$) of the Fourier plane (inset in top panel of Fig. 4d). These signals are found to oscillate with a near-180° phase difference, unambiguously indicating high-frequency beam deflection. Notably, we observe clear oscillations all the way down to a period of $\tau = 3.2$ ns ($\omega = 2\pi \cdot 0.32$ GHz). Normalizing the oscillations in APD counts to those at $\tau = 10$ μs, we find that the amplitude is approximately unaffected at $\tau = 100$ ns, and reduced by ~60% (~80%) at $\tau = 5.6$ ns (3.2 ns). Measurements of the high-frequency transmission of electrical connectors leading up to the device indicate that more than half of this reduction is in fact not due to the device itself, but rather external losses and reflections (Supplementary Note V and Supplementary Fig. S7). Reduction of pixel size and further improvement of contact quality can likely enable switching times down to a few tens of picoseconds[32,33], ultimately limited by the MoSe$_2$ mobility.

**Discussion**

While the above results were obtained at a device temperature of $T = 6$ K, almost identical data were obtained at liquid nitrogen temperatures (80 K; Supplementary Note VI and Supplementary Figure S8). Moreover, deflection was still clearly observed, although with a smaller amplitude, at a temperature easily attainable with a Peltier cooler ($T = 230$ K), and some deflection effects were even observed at room temperature. The deflection range could be further increased at all temperatures through improvements in material quality or further optimization of the hBN thicknesses and substrate permittivity. In fact, while background reflections often reduce the phase range, optimization of their phase and amplitude can increase the phase range to $2\pi$ (Supplementary Note II and Supplementary Figures S1-S3). Combined with the introduction of more pixels in next generation devices, this is expected to enable a larger ratio between the steering and beam divergence angles.

These observations demonstrate that our system can be an attractive platform for applications involving high-speed active optics, with on-chip integrability and potential for flexible transparent optics as very appealing features. In addition to the two-dimensional beam steering demonstrated here, our system can be scaled up into more advanced pixel arrays with feature sizes well below 100 nm through standard etch techniques, to enable a broad variety of other atomically thin optical elements, including atomically flat holograms with many controllable outputs and flat lenses with tunable focal length. In particular, our approach of using gates in two planes makes it possible to generate a 2D grid of $n \times n$ pixels ($2n$ independent input channels) by etching the top and bottom gates into thin perpendicular strips (see Supplementary Note VII and Supplementary Fig. S9 for further details). The use of nanoelectrode arrays is another promising upscaling route. While exfoliated flakes with areas typically exceeding 100 µm² can fit a high number of pixels, recent progress in chemical vapor deposition (CVD) growth[34] can enable scaling up to even larger devices.

Importantly, our approach also offers unique features which could unlock quantum applications that are inherently impossible with conventional spatial light modulator (SLM) technologies. In particular, quantum analogues of metasurfaces that could generate and manipulate entangled states of light have recently been proposed[35,36]. The realization of such devices relies on creating metasurfaces from locally tunable materials that can exist in a superposition of states with different reflectivity and that facilitate quantum nonlinear effects, thus calling for the exploration of new material systems for flat optics. In contrast to materials commonly used in conventional SLMs, the K/K' valley exciton species employed here has been widely shown to exhibit promising, gate-tunable quantum coherence properties and can be readily excited to a quantum superposition of states that interact only with left- and right-handed circularly polarized light. Moreover, the use of excitons confined in an atomically thin material renders it a promising approach for achieving quantum nonlinear effects mediated by exciton-exciton interactions. Recent works suggest that strong exciton-exciton interactions could be achieved in our system through reflective substrate engineering[23] or by using Rydberg exciton states[37], making active flat optics based on TMDs a very appealing platform for the investigation of quantum optical metasurfaces[35].

**Methods**
Device fabrication
In order to minimize contact resistance, crucial to high-frequency operation, we fabricated bottom contacts to the $MoSe_2$[38]. This was done by first assembling mechanically exfoliated flakes of graphene and hBN with the dry-transfer method, and placing them on a quartz substrate. After thermal annealing of the two-flake stack, platinum contacts were defined with e-beam lithography and deposited on top of the hBN flake through thermal evaporation (1 nm Cr + 19 nm Pt). The partially complete device was then thermally annealed again, before assembling mechanically exfoliated $MoSe_2$, hBN and graphene flakes and placing them on top of the contacts. Finally, extended electrical contacts to the Pt contacts and the graphite gates were deposited through thermal evaporation (10 nm Cr + 90 nm Au).

Experimental method
All measurements were conducted in a Montana Instruments cryostat, using a custom-built 4f confocal setup with a Zeiss (100x, NA=0.75, WD=4 mm) objective. Reflection spectra were measured using a halogen source and a spectrometer, and all spectra were normalized to that collected from a gold contact. Electrostatic gating was performed with Keithley 2400 multimeters for DC measurements and with an arbitrary waveform generator (Tektronix AWG710) for AC measurements. We used a Ti:Sapphire laser (M Squared) with a power of 5 µW at the sample for Fourier imaging, and imaged the reflected beam with a CMOS camera in the Fourier plane. At high frequencies, an avalanche photodetector (APD) was used to collect photons from two different parts of the Fourier plane. The time dependence was measured using a Time Correlated Single-Photon Counting system (PicoHarp 300).

Data availability

All data needed to evaluate the findings in the paper are present in the paper and the supplementary materials.

**References:**

1. Yu, N. & Capasso, F. Flat optics with designer metasurfaces. *Nature Materials* **13**, 139-150 (2014).

2. Kildishev, A. V., Boltasseva, A. & Shalaev, V. M. Planar Photonics with Metasurfaces. *Science* **339**, 1232009 (2013).

3. Lin, D., Fan, P., Hasman, E. & Brongersma, M. L. Dielectric gradient metasurface optical elements. *Science* **345**, 298-302 (2014).

4. Khorasaninejad, M. *et al.* Metalenses at visible wavelengths: Diffraction-limited focusing and subwavelength resolution imaging. *Science* **352**, 1190-1194 (2016).

5. Wang, Z. *et al.* Exciton-Enabled Meta-Optics in Two-Dimensional Transition Metal Dichalcogenides. *Nano Letters* **20**, 7964-7972 (2020).

6. Yu, N. *et al.* Light Propagation with Phase Discontinuities: Generalized Laws of Reflection and Refraction. *Science* **334**, 333-337 (2011).

7. Shaltout, A., Kildishev, A. & Shalaev, V. Time-varying metasurfaces and Lorentz non-reciprocity. *Optical Materials Express* **5**, 2459-2467 (2015).

8. High, A. A. *et al.* Visible-frequency hyperbolic metasurface. *Nature* **522**, 192-196 (2015).

9. Shaltout, A. M., Shalaev, V. M. & Brongersma, M. L. Spatiotemporal light control with active metasurfaces. *Science* **364**, eaat3100 (2019).

10. Brongersma, M. L. The road to atomically thin metasurface optics. *Nanophotonics* **10**, 643-654 (2020).

11. Wang, Y. *et al.* Electrical tuning of phase-change antennas and metasurfaces. *Nature Nanotechnology* (2021).

12. Zhang, Y. *et al.* Electrically reconfigurable non-volatile metasurface using low-loss optical phase-change material. *Nature Nanotechnology* (2021).

13  Gholipour, B., Zhang, J., MacDonald, K. F., Hewak, D. W. & Zheludev, N. I. An All-Optical, Non-volatile, Bidirectional, Phase-Change Meta-Switch. *Advanced Materials* **25**, 3050-3054 (2013).

14  Kocer, H. *et al.* Thermal tuning of infrared resonant absorbers based on hybrid gold-VO2 nanostructures. *Applied Physics Letters* **106**, 161104 (2015).

15  Belotelov, V. I. *et al.* Plasmon-mediated magneto-optical transparency. *Nature Communications* **4**, 2128 (2013).

16  Holsteen, A. L., Cihan, A. F. & Brongersma, M. L. Temporal color mixing and dynamic beam shaping with silicon metasurfaces. *Science* **365**, 257-260 (2019).

17  van de Groep, J. *et al.* Exciton resonance tuning of an atomically thin lens. *Nature Photonics* **14**, 426-430 (2020).

18  Huang, Y.-W. *et al.* Gate-Tunable Conducting Oxide Metasurfaces. *Nano Letters* **16**, 5319-5325 (2016).

19  Shirmanesh, G. K., Sokhoyan, R., Wu, P. C. & Atwater, H. A. Electro-optically Tunable Multifunctional Metasurfaces. *ACS Nano* **14**, 6912-6920 (2020).

20  Park, J. *et al.* All-solid-state spatial light modulator with independent phase and amplitude control for three-dimensional LiDAR applications. *Nature Nanotechnology* **16**, 69–76 (2020).

21  Shiue, R.-J. *et al.* Active 2D materials for on-chip nanophotonics and quantum optics. *Nanophotonics* **6**, 1329-1342 (2017).

22  Jones, A. M. *et al.* Optical generation of excitonic valley coherence in monolayer WSe2. *Nature Nanotechnology* **8**, 634-638 (2013).

23  Wild, D. S., Shahmoon, E., Yelin, S. F. & Lukin, M. D. Quantum Nonlinear Optics in Atomically Thin Materials. *Physical Review Letters* **121**, 123606 (2018).

24  Shahmoon, E., Wild, D. S., Lukin, M. D. & Yelin, S. F. Cooperative Resonances in Light Scattering from Two-Dimensional Atomic Arrays. *Physical Review Letters* **118**, 113601 (2017).

25    Back, P., Zeytinoglu, S., Ijaz, A., Kroner, M. & Imamoğlu, A. Realization of an Electrically Tunable Narrow-Bandwidth Atomically Thin Mirror Using Monolayer MoSe2. *Physical Review Letters* **120**, 037401 (2018).

26    Scuri, G. *et al.* Large Excitonic Reflectivity of Monolayer MoSe2 Encapsulated in Hexagonal Boron Nitride. *Physical Review Letters* **120**, 037402 (2018).

27    Ross, J. S. *et al.* Electrical control of neutral and charged excitons in a monolayer semiconductor. *Nature Communications* **4**, 1474 (2013).

28    Mak, K. F. *et al.* Tightly bound trions in monolayer MoS2. *Nature Materials* **12**, 207-211 (2012).

29    Yuan, H. *et al.* High-Density Carrier Accumulation in ZnO Field-Effect Transistors Gated by Electric Double Layers of Ionic Liquids. *Advanced Functional Materials* **19**, 1046-1053 (2009).

30    Sidler, M. *et al.* Fermi polaron-polaritons in charge-tunable atomically thin semiconductors. *Nature Physics* **13**, 255-261 (2016).

31    Arora, A., Nogajewski, K., Molas, M., Koperski, M. & Potemski, M. Exciton band structure in layered MoSe2: from a monolayer to the bulk limit. *Nanoscale* **7**, 20769-20775 (2015).

32    Cheng, R. *et al.* Few-layer molybdenum disulfide transistors and circuits for high-speed flexible electronics. *Nature Communications* **5**, 5143 (2014).

33    Duan, X., Wang, C., Pan, A., Yu, R. & Duan, X. Two-dimensional transition metal dichalcogenides as atomically thin semiconductors: opportunities and challenges. *Chemical Society Reviews* **44**, 8859-8876 (2015).

34    Rogers, C., Gray, D., Bogdanowicz, N. & Mabuchi, H. Laser Annealing for Radiatively Broadened MoSe2 grown by Chemical Vapor Deposition. *arXiv:1804.07880*, arXiv:1804.07880 (2018).

35    Bekenstein, R. *et al.* Quantum metasurfaces with atom arrays. *Nature Physics* **16**, 676-681 (2020).

36    Solntsev, A. S., Agarwal, G. S. & Kivshar, Y. S. Metasurfaces for quantum photonics. *Nature Photonics* **15**, 327-336 (2021).

37  Walther, V., Johne, R. & Pohl, T. Giant optical nonlinearities from Rydberg excitons in semiconductor microcavities. *Nature Communications* **9**, 1309 (2018).

38  Fallahazad, B. *et al.* Shubnikov–de Haas Oscillations of High-Mobility Holes in Monolayer and Bilayer WSe2: Landau Level Degeneracy, Effective Mass, and Negative Compressibility. *Physical Review Letters* **116**, 086601 (2016).

## Acknowledgments

We acknowledge support from the DoD Vannevar Bush Faculty Fellowship (N00014-16-1-2825 for H.P., N00014-18-1-2877 for P.K.), NSF (PHY-1506284 for H.P. and M.D.L.), NSF CUA (PHY-1125846 for S.F.Y., H.P. and M.D.L.), AFOSR MURI (FA9550-17-1-0002), ARL (W911NF1520067 for H.P. and M.D.L.), the Gordon and Betty Moore Foundation (GBMF4543 for P.K.), ONR MURI (N00014-15-1-2761 for P.K.), DOE (DE-SC0020115 for S.F.Y.), and Samsung Electronics (for P.K. and H.P.). All fabrication was performed at the Center for Nanoscale Systems (CNS), a member of the National Nanotechnology Coordinated Infrastructure Network (NNCI), which is supported by the National Science Foundation under NSF award 1541959. K.W. and T.T. acknowledge support from the Elemental Strategy Initiative conducted by the MEXT, Japan and the CREST (JPMJCR15F3), JST. A.S. acknowledges support from the Fannie and John Hertz Foundation and the Paul & Daisy Soros Fellowships for New Americans. This project has received funding from the European Union's Horizon 2020 research and innovation programme under the Marie Skłodowska-Curie grant agreement No 101023276.

## Author contributions

T.I.A., R.J.G., G.S., D.S.W., R.B., S.F.Y, P.K., H.P. and M.D.L. conceived the project. T.I.A. designed and performed the experiments with assistance from R.J.G., G.S., B.L.D., A.S., Y.Z., A.A.Z, and A.Y.J. Device fabrication was done by T.I.A., R.J.G., G.S., J.S. and X.L., and the theoretical model was developed by T.I.A., D.S.W. and R.B. T.I.A. wrote the manuscript with extensive input from the other authors. T.T. and K.W. grew the hBN crystals. S.F.Y., P.K., H.P., and M.D.L. supervised the project.

## Competing interests

Harvard University has filed a provisional patent application (No. 63/153,726) for a fast spatial light modulator based on an atomically thin reflector, with the following inventors: T.I.A., R.J.G., G.S., B.L.D., D.W., R.B., A.S., S.Y., P.K., H.P. and M.D.L.

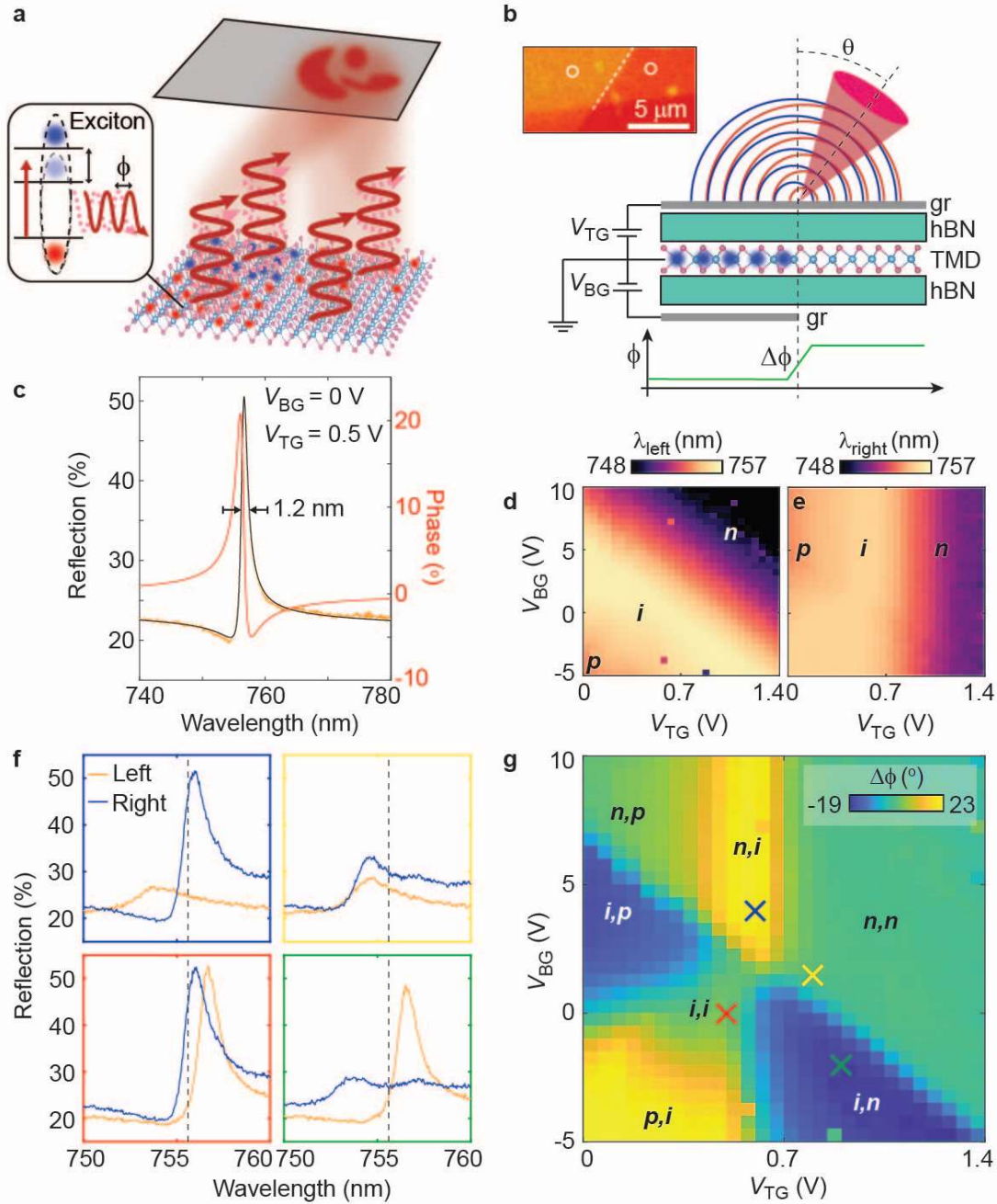

**Fig. 1: Continuously tunable phase patterning in a van der Waals heterostructure**. **a**, Schematic of our approach: patterned electrostatic doping of atomically thin transition metal dichalcogenides (TMDs) allows for spatial control of the exciton resonance (inset). Thus, a continuously tunable phase profile is imparted on the reflected wavefront, enabling wide-ranging possibilities for beam control. **b**, Schematic of SG-FET structure. Since the bottom gate only covers part of the device, the phase can be tuned independently in the two sides. The phase discontinuity in the reflected wavefront causes the two halves to constructively interfere at an angle in the far field. Inset: zoomed-in optical microscope image of the device, with gate edge indicated by white dashed line. **c**, Representative reflection spectrum (orange) from left side of gate edge in intrinsic regime ($V_{BG}$ =0 V and $V_{TG}$ =0.5 V), with asymmetric resonance fit (black), which allows for extracting the phase (red). **d,e,** Gate dependence of $\lambda_{left}$ and $\lambda_{right}$, respectively

(locations indicated by circles in inset of **b**). The exciton resonance blue-shifts upon electrostatic doping. While $\lambda_{\text{left}}$ depends on $8V_{\text{TG}}+V_{\text{BG}}$, $\lambda_{\text{right}}$ is largely independent of $V_{\text{BG}}$. The intrinsic regime appears at an offset of $V_{\text{TG}} = 0.5$ V, likely due to charge collection at the top gate. The small voltage range of the intrinsic regime suggests some doping via in-gap states. **f**, Reflection spectra from left (orange) and right (blue) side of gate edge at different gate voltage combinations shown as correspondingly colored crosses in **g**. Dashed gray line indicates $\lambda_0$=755.6 nm. **g**, Gate dependence of phase difference $\Delta\phi$ between the right and left side at $\lambda_0$=755.6 nm, computed from fits as in **c**. A tunable $\Delta\phi$-range of 42° is achieved. Large positive $\Delta\phi$ is achieved when $\lambda_{\text{left}} < \lambda_0 < \lambda_{\text{right}}$ (blue cross), while large negative $\Delta\phi$ is achieved when $\lambda_{\text{right}} < \lambda_0 < \lambda_{\text{left}}$ (green). $\Delta\phi$ is closer to zero when either both sides are doped ($\lambda_{\text{left}}, \lambda_{\text{right}} < \lambda_0$; yellow) or both are intrinsic ($\lambda_0 < \lambda_{\text{left}}, \lambda_{\text{right}}$; red).

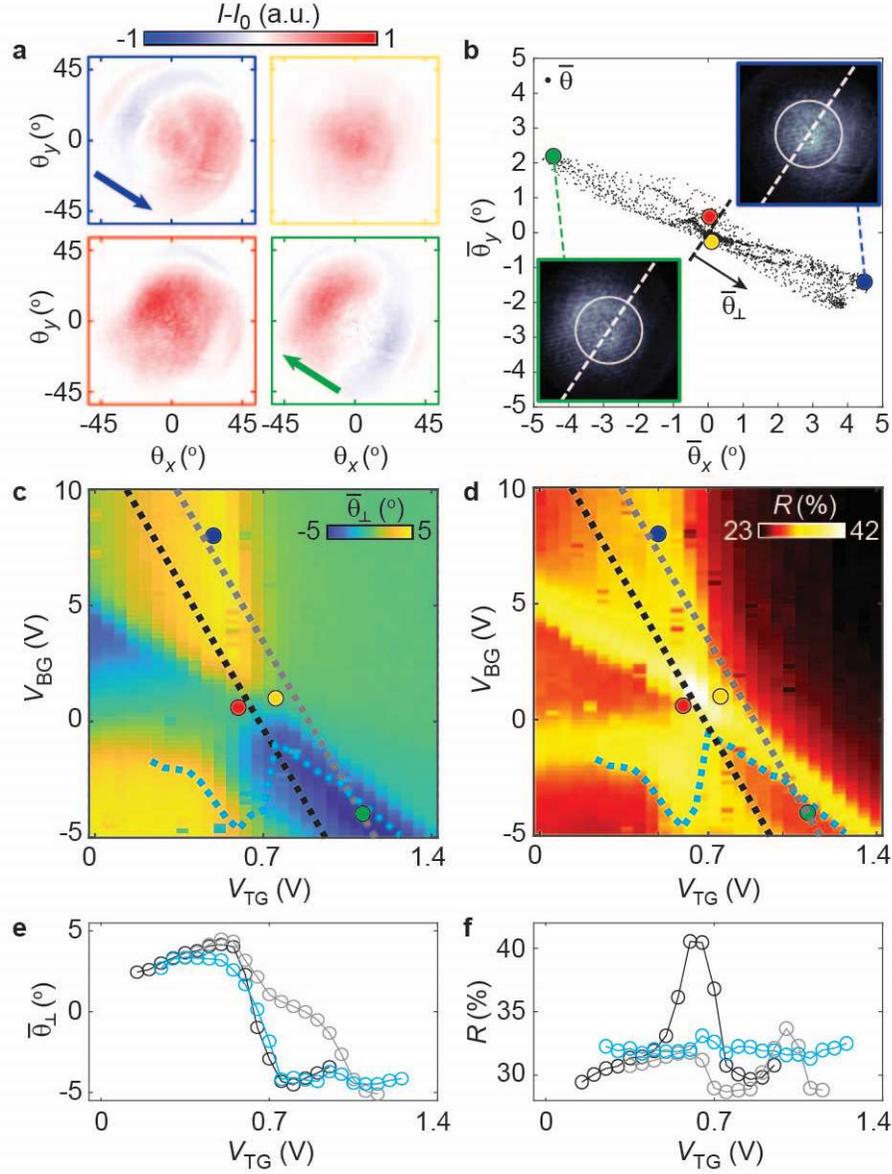

**Fig. 2: Continuously tunable beam steering. a**, Fourier images of reflected beam ($\lambda_0$=755.6 nm) in the four regimes after subtracting the reflection in the highly doped regime ($V_{BG}$ =10 V and $V_{TG}$ =1.4 V). When the exciton resonance is blue-shifted past $\lambda_0$ in only one side of the device (blue, green), the beam is deflected away from that side. If neither or both are blue-shifted past $\lambda_0$, the phase difference is small and little deflection is observed (red, yellow). **b**, Scatter plot of the beam deflection ($\bar{\theta}_x, \bar{\theta}_y$) for the full range of gate voltages, showing that the deflection is perpendicular to the gate edge (dashed line) and continuously tunable. Inset: Fourier images without background subtraction. **c**, The gate dependence of the deflection perpendicular to the gate edge ($\bar{\theta}_\perp$) is in very good agreement with that of the phase difference shown in Fig. 1g. **d**, Gate-dependence of reflection amplitude. Regions with high reflection indicate that one of the resonances crosses through $\lambda_0$. **e,f**, Linecuts indicated by dashed black, teal and gray lines in **c** and **d**, respectively, highlighting the continuous steering capability and that the reflection can be kept relatively constant while deflecting the beam. Connecting lines in **e** and **f** are guides to the eye.

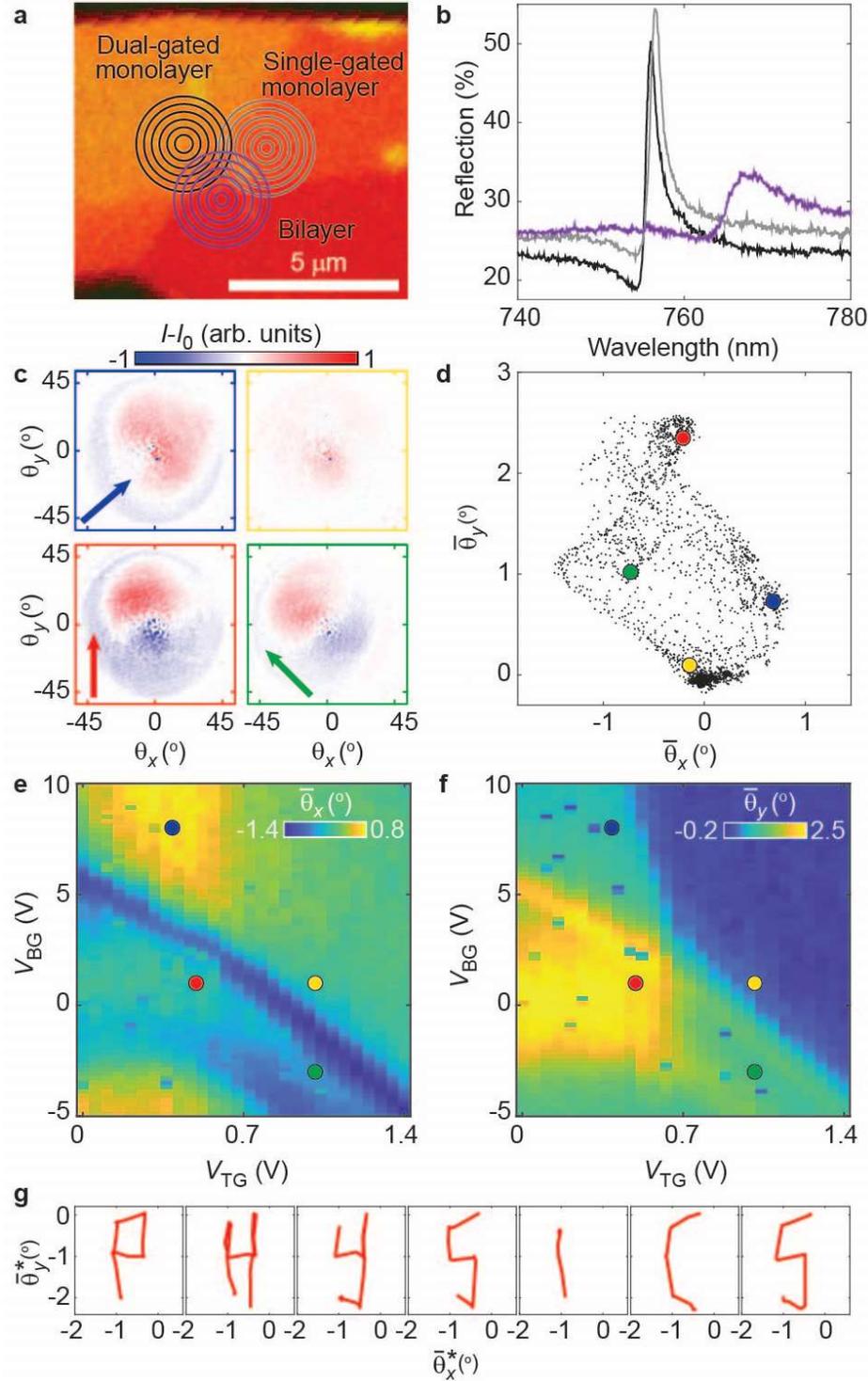

**Fig. 3: Two-dimensional beam steering. a**, Zoomed-in optical image of device, indicating regions of monolayer and bilayer MoSe$_2$, as well as gate coverage. By tuning the relative phase and amplitude of the reflection from the three regions, more intricate wavefront phase profiles can be achieved. **b**, Reflection spectra from the dual-gated monolayer (black), single-gated monolayer (gray) and the bilayer (purple) in the intrinsic regime ($V_{BG}$ = 0 V and $V_{TG}$ = 0.5 V). Since the resonance is red-shifted in the bilayer, it acts as a dielectric (non-resonant) reflector at the wavelengths used here. **c**, Fourier images of reflected beam

($\lambda_0$=754.1 nm) in the four regimes after subtracting reflection in the highly doped regime ($V_{BG}$ =10 V and $V_{TG}$ =1.4 V). The beam is now steered in two dimensions. Red: when both monolayer resonances are red-shifted relative to $\lambda_0$, their phase is higher than in the bilayer region, causing the beam to deflect upwards. Blue and green: when one of the monolayer resonances is kept red-shifted relative to $\lambda_0$, the beam is deflected towards that monolayer region. **d**, Scatter plot of the center-of-mass deflection ($\bar{\theta}_x, \bar{\theta}_y$), with the points from **c** highlighted. The set of beam deflections now span a two-dimensional area. **e,f**, Gate dependence of $\bar{\theta}_x$ (**e**) and $\bar{\theta}_y$ (**f**). While the gate dependence resembles that of the phase in Fig. 1g, $\bar{\theta}_x$ and $\bar{\theta}_y$ are now less coupled. **g,** Center-of-mass deflection tracing out "PHYSICS" (rotated 148° counter-clockwise) by applying a sequence of gate voltage combinations.

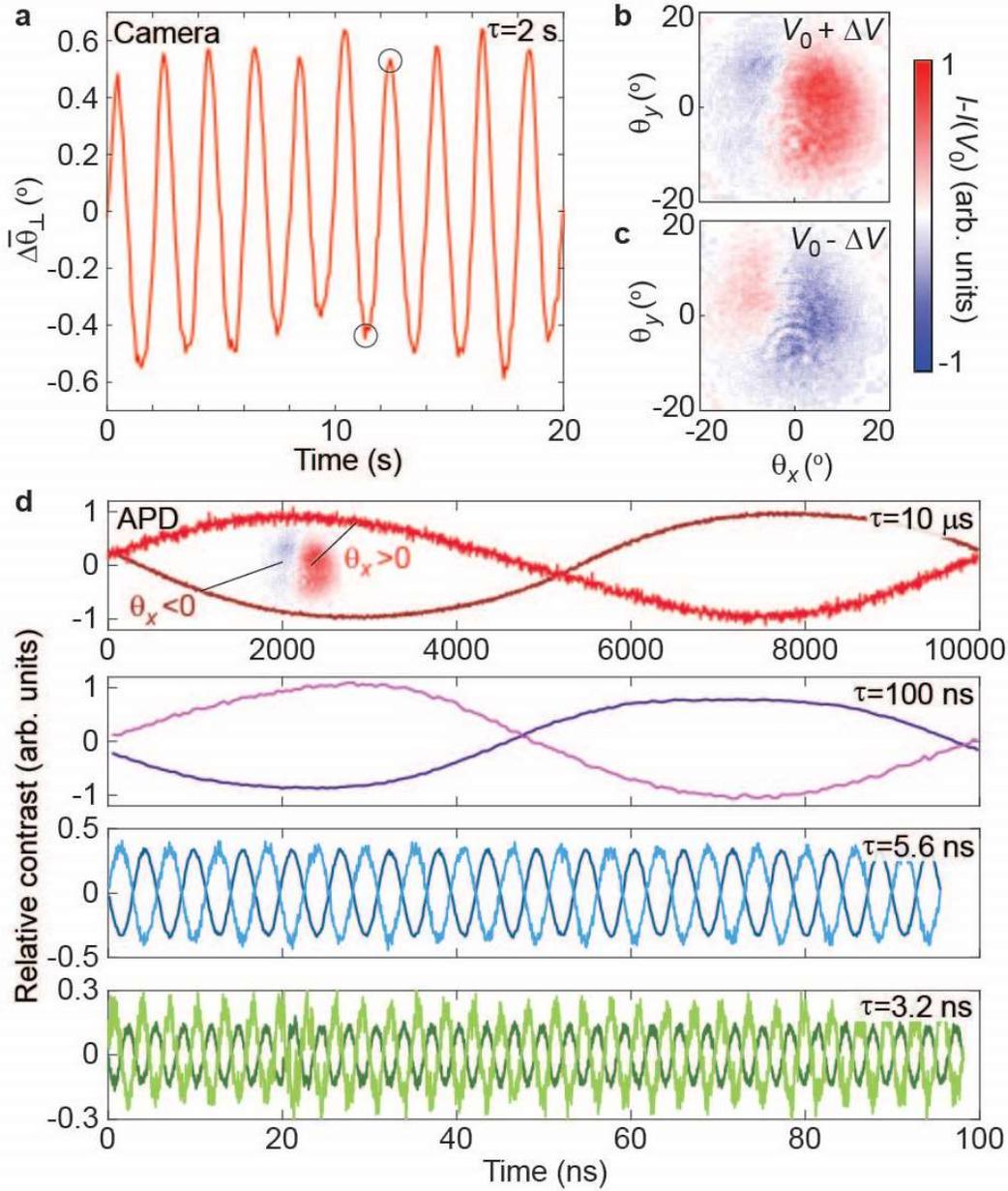

**Fig. 4: High-frequency beam steering. a**, Oscillations of center-of-mass deflection ($\lambda_0 = 754.5$ nm) induced by oscillating back gate voltage (offset, $V_0 = 0.7$ V; amplitude, $\Delta V = 0.45$ V; period, $\tau = 2$ s) and $V_{TG} = 0.64$ V. **b,c**, Fourier plane images collected at $V_{BG} = V_0 + \Delta V$ (**b**) and $V_{BG} = V_0 - \Delta V$ (**c**), as indicated by circles in **a**, after subtracting reflection at $V_{BG} = V_0$. An inverted telescope was used to shrink the beam to simplify the subsequent APD measurements. **d**, Photon count oscillations measured with APD at $\tau = 10$ μs, 100 ns, 5.6 ns and 3.2 ns (top to bottom). Darker (lighter) shade curves show photon counts from left (right) side of Fourier plane, as indicated by the inset in the top panel. All curves are normalized to the corresponding contrast at $\tau = 10$ μs.

Supplementary Materials for

# Beam steering at the nanosecond time scale with an atomically thin reflector

Trond I. Andersen[1], Ryan J. Gelly[1], Giovanni Scuri[1], Bo L. Dwyer[1], Dominik S. Wild[1,2], Rivka Bekenstein[1,3], Andrey Sushko[1], Jiho Sung[1,4], You Zhou[1,4,5], Alexander A. Zibrov[1], Xiaoling Liu[1], Andrew Y. Joe[1], Kenji Watanabe[6], Takashi Taniguchi[7], Susanne F. Yelin[1], Philip Kim[1,8], Hongkun Park[1,4,†], Mikhail D. Lukin[1,†]

[1]Department of Physics, Harvard University, Cambridge, MA 02138, USA
[2]Max Planck Institute of Quantum Optics, Hans-Kopfermann-Straße 1, D-85748 Garching, Germany
[3]ITAMP, Harvard-Smithsonian Center for Astrophysics, Cambridge, MA 02138, USA
[4]Department of Chemistry and Chemical Biology, Harvard University, Cambridge, MA 02138, USA
[5]Department of Materials Science and Engineering, University of Maryland, College Park, MD, USA
[6]Research Center for Functional Materials, National Institute for Materials Science, 1-1 Namiki, Tsukuba 305-0044, Japan
[7]International Center for Materials Nanoarchitectonics, National Institute for Materials Science, 1-1 Namiki, Tsukuba 305-0044, Japan
[8]John A. Paulson School of Engineering and Applied Sciences, Harvard University, Cambridge, MA 02138, USA
[†]To whom correspondence should be addressed: lukin@physics.harvard.edu and hongkun_park@harvard.edu

**Contents:**

Supplementary Notes I-VIII
Supplementary Figs. S1 to S13

**Other Supplementary Materials for this manuscript include the following:**

Supplementary Video S1

**Supplementary Note I.  Phase extraction through spectrum fitting**

The interference between the reflection from the excitons and from other interfaces within our system causes an asymmetric line shape. The resultant reflection arises from the infinite sum of optical paths created by different combinations of reflection and transmission at the various interfaces. By assuming that the background reflection in the absence of excitonic response is constant ($|r_0|e^{i\phi_0}$) in the small spectral range considered in our fits, the reflection can be written on the form[1]:

$$r(\omega) = \frac{\gamma_0 \left(\frac{2\omega_0 G}{c}\right)^2}{\omega - \omega_0 + \frac{i\Gamma}{2}} + r_0 = \frac{A + iB}{\omega - \omega_0 + \frac{i\Gamma}{2}} + r_0, \quad (1)$$

where $\gamma_0$ is the free-space radiative rate of the excitons, $\Gamma$ is the total linewidth, and $\omega_0$ is the exciton resonance frequency. $G$ is the Green's function that propagates fields from the TMD to the far-field, and incorporates the effects of the background reflectors. We therefore stress that the first term is not equivalent to the reflection of a freestanding TMD. Because $G$ is generally complex, we replace the numerator of the first term with a complex number $A + iB$, where $A$ and $B$ are real.

Introducing $\delta = \omega - \omega_0$, we find expressions for the amplitude and phase of the reflection:

$$R(\omega) = |r(\omega)|^2 = |r_0|^2 + \frac{A^2 + B^2 + B\Gamma|r_0| + 2A|r_0|\delta}{\delta^2 + \frac{\Gamma^2}{4}} \quad (2)$$

$$\phi(\omega) - \phi_0 = \tan^{-1}\left(\frac{B\delta - \frac{A\Gamma}{2}}{A\delta + \frac{B\Gamma}{2} + |r_0|(\delta^2 + \frac{\Gamma^2}{4})}\right) \quad (3)$$

By fitting the obtained reflection spectra, we extract the parameters $|r_0|$, $A$, $B$, $\delta$ and $\Gamma$, and can thus compute the phase. Since we are only concerned with relative phases in our work, we set $\phi_0 = 0$.

**Supplementary Note II.  Effects of background reflections**

Interestingly, while the background reflections typically reduce the phase range to a value below $\pi$ (range of the pure exciton resonance), they can actually also increase the phase range to $2\pi$ if their phase and amplitude are chosen correctly. Hence, the phase range is not inherently limited in our approach. To understand how this is possible, we write the total reflection as:

$$r(\omega) = \frac{C}{2(\omega - \omega_0)/\Gamma + i} + r_0,$$

where $\Gamma$ is the total exciton linewidth, $\omega_0$ is the exciton resonance frequency and $r_0 = |r_0|e^{i\phi}$ is the background reflection that would be observed in the absence of excitons. We note that $C$ is also complex. Crucially, while the phase of the first term only varies by $\pi$ across the resonance, the phase of the full sum can vary by $2\pi$. This can be intuitively understood by considering the development of $r(\omega)$ in the complex plane as shown in Fig. S1c: as $\omega$ is moved past the resonance, the reflection traces out a circle with radius $|C|/2$ centered at $r_0 - iC/2$. (In reality, the slight frequency dependence of the background reflection causes a small deformation of this circular path). Hence, in order to achieve the desired $2\pi$ phase range, this circle needs to encompass the origin, which is the case so long as $\left|r_0 - \frac{iC}{2}\right| < |C|/2$.

To support this theoretical picture, we show reflection spectra from additional, ungated devices demonstrating the full $2\pi$ phase range across the exciton resonance (Fig. S1), through more optimized background reflections and stronger TMD reflection in those devices. As can be seen from Fig. S1c, the reflection traces encompass the origin in the complex plane, indicating the full $2\pi$ phase range (also evident in Fig. S1b). However, further optimization is required to make the circle more centered at the origin to prevent large variations in reflection amplitude, as can be seen in Fig. S1a.

In order to understand how the effects of background reflections depend on hBN thickness and substrate, we conducted a detailed quantitative analysis of the reflections within our device structure, using the transfer matrix method. This theoretical model is described in further detail in Ref. [1], and reproduces our experimental observations very well, as can be seen in Fig. S1a. We note that our model accounts for not only the effect of hBN thickness on $r_0$, but also on $C$, which depends on the Green's function that propagates fields through the heterostructure. Fig. S2e-h shows the calculated phase range for a wide range of hBN thicknesses and several substrate types (Fig. S2a-d). Crucially, our analysis demonstrates that a phase range of $2\pi$ can be achieved in a substantial part of the parameter space. We also show the maximum reflection amplitude in Fig. S2i-l, and the relative amplitude variations in Fig. S2m-p. Achieving optimal operation also requires simultaneously maximizing the reflection amplitude and minimizing its variation across the resonance. Fig. S2 shows that the use of a gold reflector under the device (Fig. S2c,d) can be a promising way to achieve desirable values for all the three criteria simultaneously.

Interestingly, as shown in the rightmost inset of Fig. S1c, the background reflections can in principle be optimized such that $r_0 = \frac{iC}{2}$, causing the reflection trace to form a circle that is centered at the origin in the complex plane. Notably, in this case, the reflection amplitude stays constant across the whole exciton resonance. Thus, the exciton-based approach shown in our work could in principle allow for making pixels with both the full phase range of $2\pi$ and constant reflection. For instance, Fig. S3 shows the numerically calculated reflection of a device placed on a gold covered substrate with top and bottom hBN-thicknesses of 28 and 40 nm, respectively. (Note that a thin hBN layer on top of the gold as shown in Fig. S2c would prevent shorting bottom graphene gates.) To here evaluate the optimal case, we have used the exciton properties of device C, which has a stronger reflection than devices A and B. (Note that this is not necessary to achieve constant reflection amplitude but increases the constant reflection amplitude). Notably, we find a full phase range of $2\pi$, reflection in excess of 84%, and near constant (<1% variation) reflection amplitude. We note that the optimal hBN thicknesses and choice of substrate depend on the exact radiative and non-radiative rates of the excitons, which vary from device to device. We therefore emphasize that while our quantitative analysis provides a deeper understanding of the effects of background reflections and demonstrates that our system is not inherently limited with regards to phase range and reflection amplitude, further studies are required to experimentally achieve complete optimization. A promising avenue for this task is utilizing "feedback fabrication" schemes where additional hBN layers are added after characterizing the TMD. Another appealing path involves the use of a substrate with globally tunable refraction index. It should be stressed that the refractive index of such a substrate would only need to be fine-tuned once and only on a global level, and would therefore not require the fast, local tunability demonstrated with a TMD in our work.

**Supplementary Note III.    Gate edge localization**
In order to position the laser beam at the gate edge, we first locate the edge by sweeping the galvanometric mirrors while measuring the reflection using a broadband halogen lamp. To further optimize the alignment of the laser spot with the gate edge, we measure beam steering in a few locations near the gate edge (example shown in Fig. S4). The relative contributions from the two sides vary as the spot is moved across the edge, and we balance the contributions by finding the location with the largest and most symmetric deflection range. Far away from the gate edge, only small deflections (in arbitrary directions) are observed (Fig. S5a,b), likely due to local inhomogeneity, causing small phase gradients that change slightly with doping. The non-zero width of the deflection path at the gate edge is likely caused by such inhomogeneity. Independent control of different parts of the wavefront, essential to amplitude stabilization (Fig. 2f in main text), two-dimensional steering and further upscaling possibilities, is of course only possible at the split-gate edge.

**Supplementary Note IV.    Theoretical modeling of beam deflection**
We here present a theoretical model of the beam deflection, the predictions of which are in very good agreement with the experimental observations presented in the main text. Defining the gate edge to lie along $x = 0$, we model the reflection profile of our system as:

$$r(\omega, \boldsymbol{r}) = \begin{cases} r_{\text{tot,L}}(\omega)\delta(z), & x < 0 \\ r_{\text{tot,R}}(\omega)\delta(z), & x > 0, \end{cases} \quad (4)$$

where $r_{\text{tot,L(R)}}(\omega)$ is the (gate-dependent) combined spectrum of the exciton and background reflections on the left (right) side of the gate edge. The combined spectrum has a smaller available phase range than that of the exciton resonance itself (0-180°) due to the interference with the background reflections. While the background reflections come from multiple interfaces both above and below the TMD, the combined system can be modeled as an equivalent reflector in the plane of the TMD ($z = 0$) by including the additional phases due to the *z*-displacements in $r(\omega)$. The incoming field is given by the two-dimensional Gaussian distribution $E(\rho) = E_0 \exp(-\rho^2/4\sigma^2)$, where $\rho$ is the radial coordinate in the plane of the TMD and $\sigma$ is the standard deviation of the incoming intensity distribution, $I_{\text{in}}(\rho) = c\varepsilon|E(\rho)|^2$. For a diffraction limited spot, one finds $\sigma = 0.42\lambda/(2\text{NA})$, where *NA* is the numerical aperture of the objective[2]. The reflected intensity at a position $\boldsymbol{r}$ is then given by:

$$I(\boldsymbol{r}, \omega) = c\varepsilon \left| \int r(\omega, \boldsymbol{r}') E(\boldsymbol{r}') \frac{\exp(ik|\boldsymbol{r} - \boldsymbol{r}'|)}{\lambda|\boldsymbol{r} - \boldsymbol{r}'|} d\boldsymbol{r}' \right|^2, \quad (5)$$

Computing the integral for the far-field ($r \to \infty$) at a polar angle $\theta$, and azimuthal angle $u$ ($u=0$ is perpendicular to the gate edge), one finds:

$$I(\theta, u, \omega) \propto \exp(-2F^2(\theta)) \left| r_{tot,R}(\omega)[1 - i\,\mathrm{erfi}(F(\theta) \cdot \cos u)] + r_{tot,L}(\omega) \cdot [1 + i\,\mathrm{erfi}(F(\theta) \cdot \cos u)] \right|^2, \quad (6)$$

where $F(\theta) = k\sigma \cdot \sin\theta$, and $\mathrm{erfi}(x)$ is the imaginary error function. Thus, we find that the reflections from the two sides interfere constructively for $\theta, u$ that satisfy:

$$\theta = \sin^{-1}\left(\frac{\mathrm{erfi}^{-1}(\tan \Delta\phi/2)}{k\sigma \cos u}\right), \quad (7)$$

where $\Delta\phi$ is the phase difference between the reflections from the two sides. In the small phase difference limit, this simplifies to $\theta = \Delta\phi \cdot \sqrt{\pi}/(4k\sigma \cos u)$, equivalent to two localized sources at $x = \pm 2\sigma/\sqrt{\pi}$. Due to the factor $\exp(-2F^2(\theta))$, the intensity maximum appears at a somewhat different angle. In cases where the amplitudes of the two reflections can be approximated to be the same, the integral simplifies to (in the small $\Delta\phi$ limit):

$$I(\theta, u, \omega) \propto 4e^{-2F^2(\theta)} |r_{tot}(\omega)|^2 (1 + \mathrm{erfi}(F(\theta) \cdot \cos u) \cdot \Delta\phi), \quad (8)$$

Since the maximum intensity appears at small $\theta$ in the small $\Delta\phi$ limit, and $\mathrm{erfi}(x) \sim 2x/\sqrt{\pi}$ for small $x$, we find:

$$(u, \theta)_{max} = \left(0, \frac{\Delta\phi}{2\sqrt{\pi}k\sigma}\right), \quad (9)$$

Finally, the center-of-mass deflection perpendicular to the gate edge is found by using the Taylor expansion of $\mathrm{erfi}(x)$, since the integral requires evaluating $\mathrm{erfi}(x)$ up to $x = k\sigma \sin(\theta_c)$ where $\theta_c$ is the collection angle:

$$\boxed{\bar{\theta}_\perp = \bar{\theta}_x = \frac{\int I\,\theta \cos(u)\,d\Omega}{\int I\,d\Omega} = \frac{\Delta\phi}{\sqrt{2\pi}k\sigma} = \frac{\Delta\phi \cdot \mathrm{NA}}{0.42 \cdot \sqrt{2\pi^3}}} \quad (10)$$

In solving the integrals, we have made the approximation $\sin\theta \sim \theta$, which is an acceptable approximation when $k\sigma > 1$, since the exponential factor then suppresses terms at large $\theta$. For a diffraction-limited spot, one finds $k\sigma \sin\theta_c = k\sigma \cdot \mathrm{NA} = 1.3$, or $e^{-2(k\sigma \sin\theta_c)^2} = 0.03$.

We compare eqn. (10) with the exact (numerically solved) deflection in Fig. S6 for NA=0.75 ($k\sigma \sim 1.8$), and find that they are very similar. Moreover, the theoretically predicted deflection range of $\sim 10°$ ($\pm 5°$) for a phase difference range of $42°$ is in excellent agreement with our experimental observations.

**Supplementary Note V.      High-frequency transmission characterization**
The high-frequency measurements presented in Fig. 4d in the main text show clear beam steering down to switching times of 1.6 ns ($\omega = 2\pi \cdot 316$ MHz). However, the amplitude is found to decrease at the highest frequencies, which could be due to either the *RC*-time of the device itself or the external cabling leading up to the device. In order to determine the impact of the latter, we use a Vector Network Analyzer (VNA) to characterize the high-frequency performance of parts of the external cabling that was used in the measurement (Fig. S7). While we cannot recreate the full circuit in the absence of the device without introducing additional components, this measurement places an upper bound on the transmission of the full cabling. We find a substantial reduction in the transmission at the higher frequencies used in our work, with S12 parameters of 50% and 34% at $\omega = 2\pi \cdot 178$ MHz and $\omega = 2\pi \cdot 316$ MHz, respectively. This gives a lower bound on the decay factor due to the device itself of 75% and 61% at the two frequencies, respectively.

**Supplementary Note VI.      Temperature dependence**
Fig. S8a shows reflection spectra obtained in the right side of the device in the intrinsic regime at five different temperatures ranging from 6 K to 300 K. Consistent with previous studies[1], we find that the exciton resonance broadens, decreases in amplitude, and red-shifts with increasing temperature. While the phase of the exciton resonance still changes from 0° to 180° across the resonance, the decrease in amplitude causes a reduction in the available phase range of the combined background and exciton signal (Fig. S8b). However, we find that the reflection amplitude and phase range remain almost the same at liquid nitrogen temperatures (80 K) as at 6 K, and the exciton resonance is still clearly visible at higher temperatures.

Since the resonance wavelength changes with temperature, we conduct gate-dependent beam steering measurements for a range of laser wavelengths at each temperature to ensure proper comparison across temperatures (Fig. S8c), and present scatter plots from the optimal wavelengths in Fig. S8d. The deflection ranges shown in Fig. S8c are based on the set of gate voltage combinations used in the main text. Consistent with the reduction in phase range, the beam deflection range decreases with temperature (Fig. S8c,d). Nevertheless, the deflection range remains almost 8° at 80 K and 2° at 150 K. Some deflection of the full reflection can also be observed at 230 K (-43 °C) and even at room temperature (Fig. S8d). As discussed in Supplementary Note II, the operation at these higher temperatures could be improved by optimizing the background reflections, thus allowing for a larger phase range of the combined reflection.

**Supplementary Note VII.      Upscaling to more pixels**
We here discuss two realistic avenues for upscaling: first, through standard etching techniques, the gates can be divided into very many pixels with sub-100 nm dimensions and be employed in the same way as demonstrated in our current work. One appealing design enabled by our use of gates in two different planes is to etch the top and bottom gates into long, perpendicular strips (Fig. S9). This way, the gate strips form a two-dimensional grid of pixels, where the phase of pixel (*i,j*) is set by the voltages applied to top gate strip *i* and bottom gate strip *j*, through the same double capacitor mechanism as in our current work. While such a design enables independent control of $2n$ degrees of freedom for an $n \times n$-grid, applications that would require more independent channels could be achieved through further reductions in the pitch of nanoelectrode arrays. By placing our device structure on top of such arrays, full independent control of all pixels could be achieved. Since the mechanism behind such control would still be electrostatic doping, the switching rates are expected to be similar to the ones demonstrated in our current work. Through scaling up the device to more pixels, more complex beam shaping can be achieved, and a larger incident beam spot (smaller beam divergence angle) can be used.

**Supplementary Note VIII.    Polarization dependence**
In order to test how the performance of our beam steering device depends on the polarization of the incoming light, we measure the amplitude and deflection of the reflection using four different (linear) polarization angles (Fig. S10). To remove any effects of polarization-dependent optical components in the beam path (e.g. beam splitters), we normalize the amplitude of the reflection to that without excitonic effects, obtained by heavily doping the device. Since the latter is not affected by gradients in exciton properties, we assume it to be an isotropic reference. Fig. S10a-d shows the gate dependence of the integrated reflection, indicating no systematic variations with polarization angle. Plotting the gate dependence of the deflection angle (Fig. S10e-h) and a scatter plot of all deflections (Fig. S10i-l), we also observe no systematic polarization dependence of the deflection range or the deflection direction. The small (non-systematic) variations observed in Fig. S10 are expected to be due to small changes in spot location as the polarization very weakly affects the beam path. The robustness to polarization variations stands in contrast to many other types of beam steering devices and is a very useful feature for many applications.

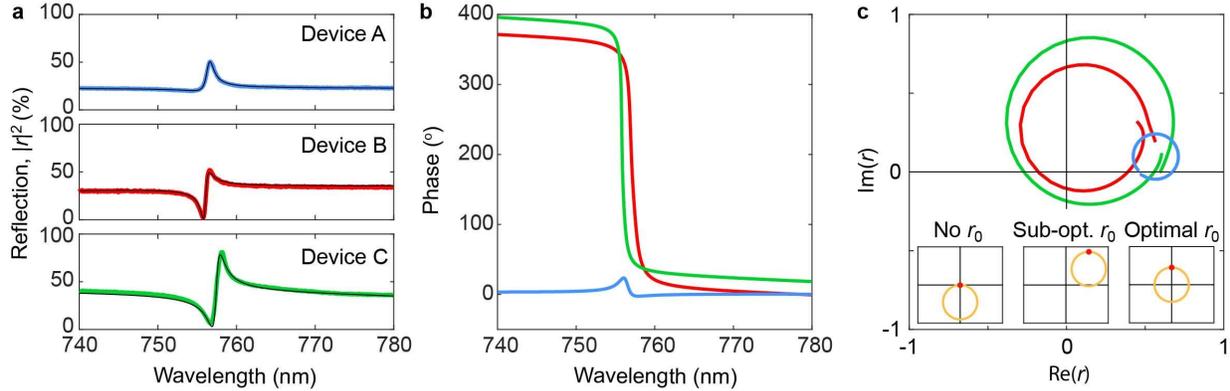

**Fig. S1: Demonstration that $2\pi$ phase range is possible in the exciton-based approach. a,** Reflection spectra (colored curves) with fits based on the transfer matrix method (black) for devices A, B and C (top to bottom). Top/bottom hBN thicknesses and substrates for the three devices are, A: 8 nm/60 nm, quartz; B: 70 nm/100 nm, Si/SiO$_2$; C: 55 nm/86 nm, Si/SiO$_2$. **b,** Phase extracted from fits in **a**, demonstrating phase range of $2\pi$ across exciton resonance in devices B and C. **c,** Complex plane representation of reflection parametrized by wavelength, demonstrating that the reflection traces a circle around the origin in devices B and C. Insets: Illustrations of complex $r$ in cases with zero, sub-optimal and optimal background reflection, giving phase ranges of $\pi$, $<\pi$ and $2\pi$, respectively. Red dot indicates $r_0$. Without loss of generality, the parameter $C$ is set to be real.

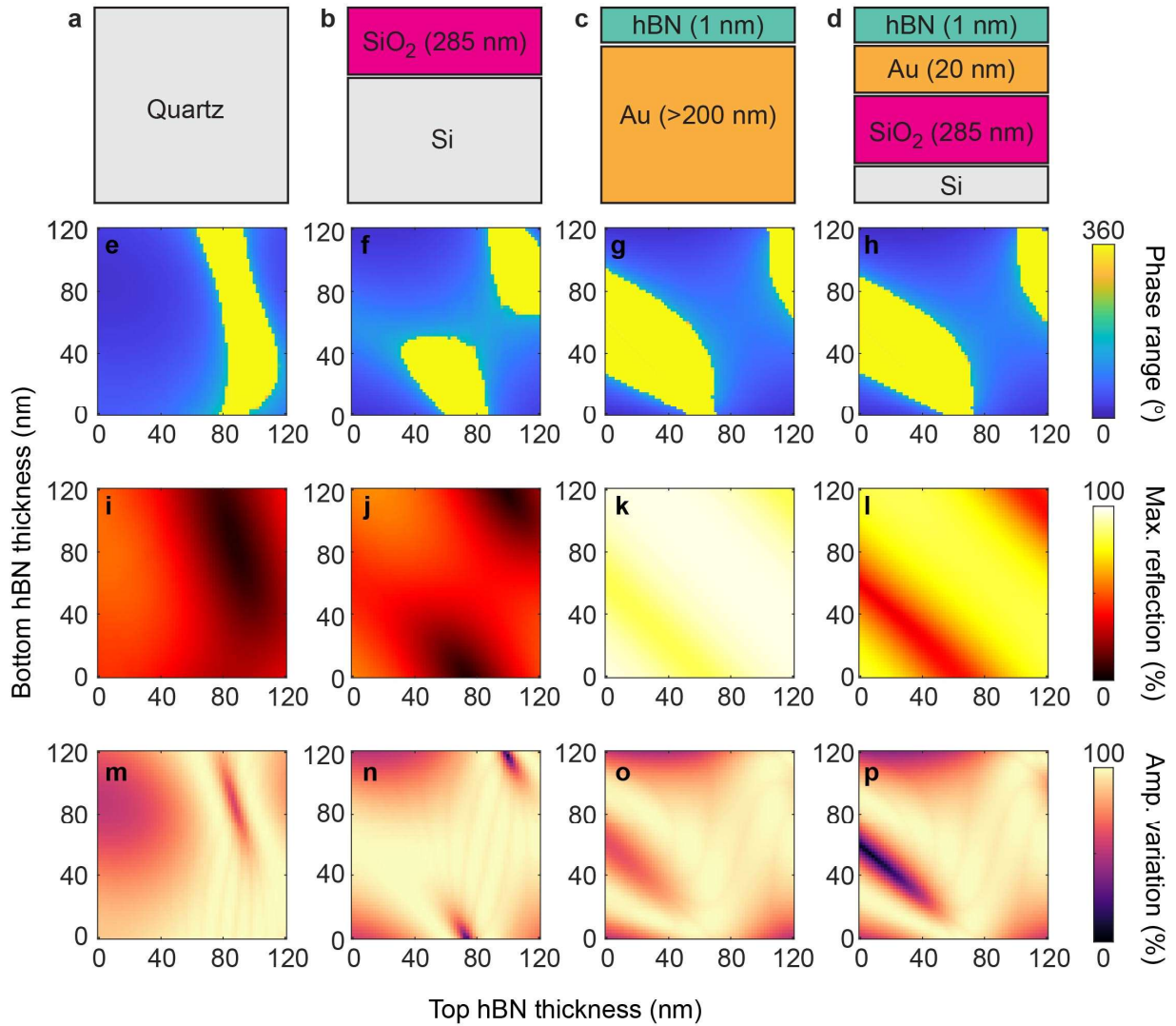

**Fig. S2: Effect of hBN thickness and substrate on phase range, reflection and amplitude stability. a-d,** Schematics of different substrate options. A thin (1 nm) hBN layer is included in **c** and **d** since the gold would otherwise short the bottom graphene gates. **e-h,** Phase range as a function of hBN thicknesses for the four substrate options, demonstrating that a phase range of $2\pi$ is achievable in a substantial part of the parameter space. **i-l,** Maximum reflection amplitude. **m-p,** Relative amplitude variation. Simultaneous optimization of all three characteristics could be achieved using substrates that include a gold layer (**c,d**).

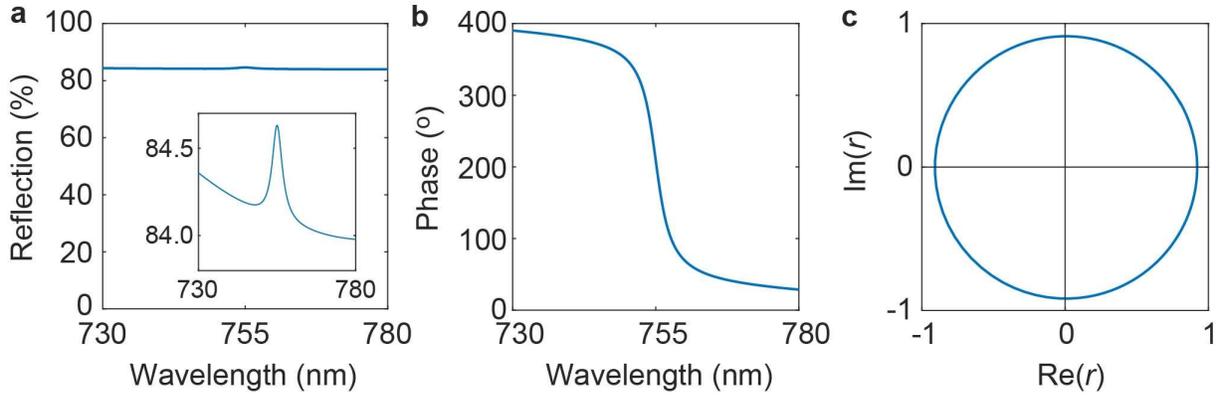

**Fig. S3: Performance with optimized device design. a,** Numerically calculated reflection spectrum of device on gold-covered substrate with top and bottom hBN thicknesses of 28 and 40 nm, respectively. A large (∼ 84%) reflection amplitude is achieved while keeping the variation smaller than 1% in this optimized design. Inset: Zoomed in version of the main plot. **b,** Phase of reflection, showing $2\pi$ range across the resonance. **c,** Complex plane representation of reflection parametrized by the wavelength, demonstrating that the circle is very close to being centered at the origin.

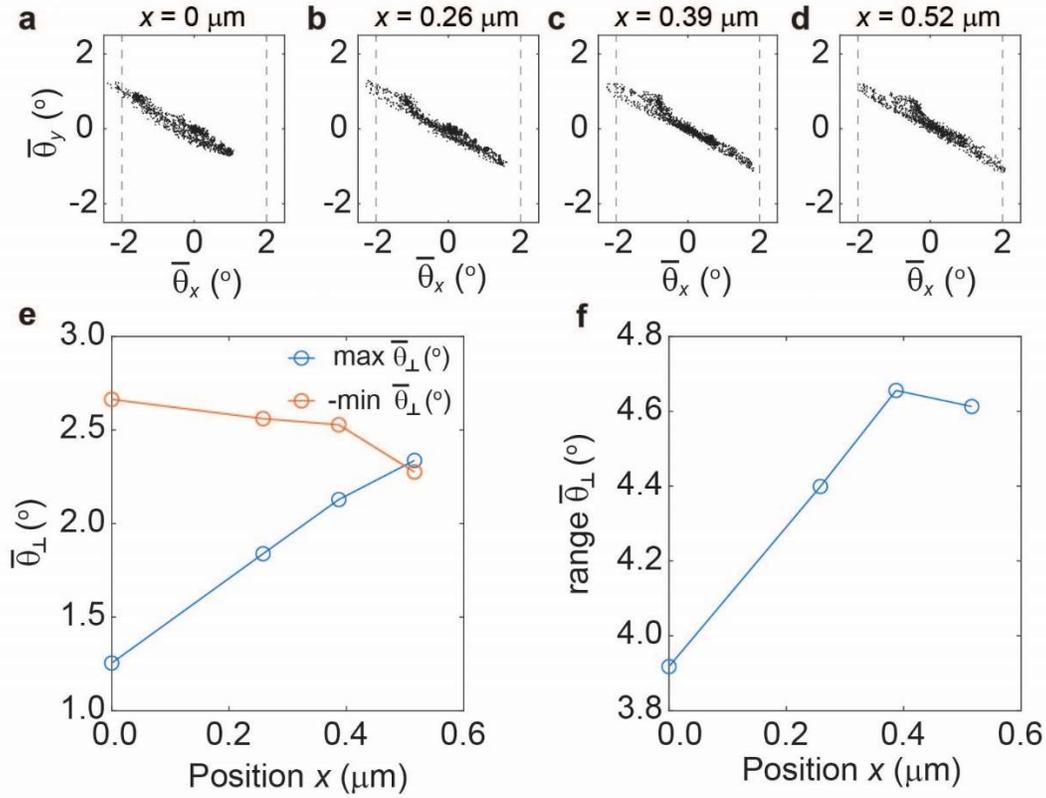

**Fig. S4: Example of gate edge localization. a-d**, Scatter plot of center-of-mass deflection ($\bar{\theta}_x$, $\bar{\theta}_y$) for the same set of gate voltages as in the main text in four different positions along a line crossing the gate edge ($\lambda_0$=758.4 nm, $T$=80 K). ($x$=0 is arbitrarily defined). **e,f**, Position dependence of the extrema (**e**) and range (**f**) of the beam deflection perpendicular to the gate edge. The contributions from the two sides vary across the gate edge, and are balanced by making the deflection range as large and symmetric as possible (around $x$=0.4 μm).

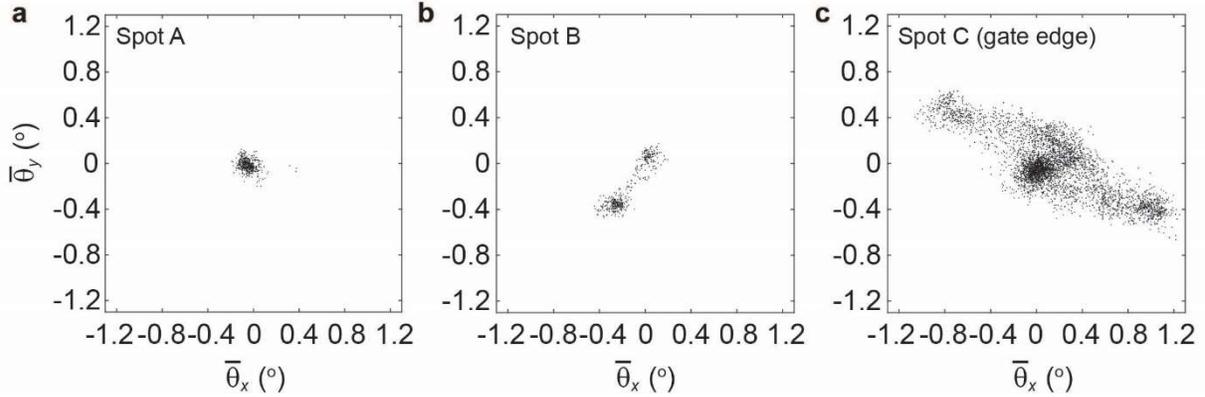

**Fig. S5: Deflection behavior away from gate edge. a,b**, Scatter plot of center-of-mass deflection ($\lambda_0$=754.5 nm) in two spots that are 2 µm to the left and right of the gate edge, respectively. All doping regimes are covered by sweeping the full range of the (global) top gate (0 V < $V_{TG}$ < 1.4 V). The deflection is found to be very small far away from the gate edge, with only small deviations in arbitrary directions (example shown in **b**), likely caused by inhomogeneity. Inhomogeneity is likely the cause of the non-zero width of the deflection path at the gate edge. **c**, Scatter plot of center-of-mass deflection ($\lambda_0$=754.5 nm) with the beam spot centered at the gate edge, for the same set of gate voltages as in the main text. Measurements in **a**-**c** are performed with ~ 2.5x enlarged spot-size, as in Fig. S11b.

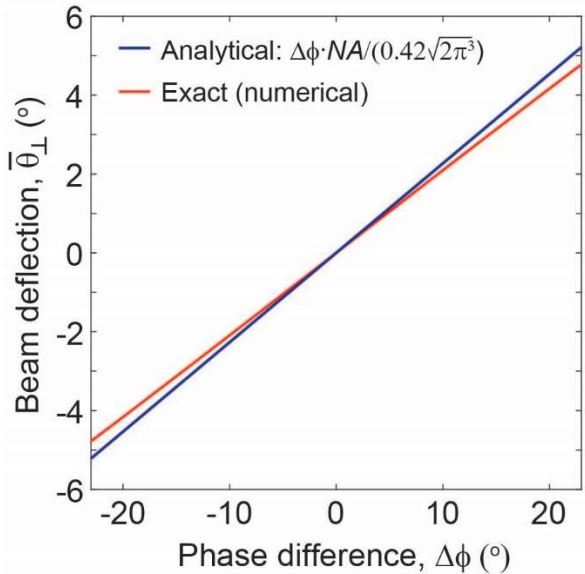

**Fig. S6: Theoretically predicted beam deflection.** Analytically (blue) and numerically (red) predicted beam deflection as a function of phase difference between the two sides of the gate edge. The predicted beam deflection range of approximately 10° ($\pm$5°) for a phase range of 42° is in excellent agreement with the experimental results presented in the main text.

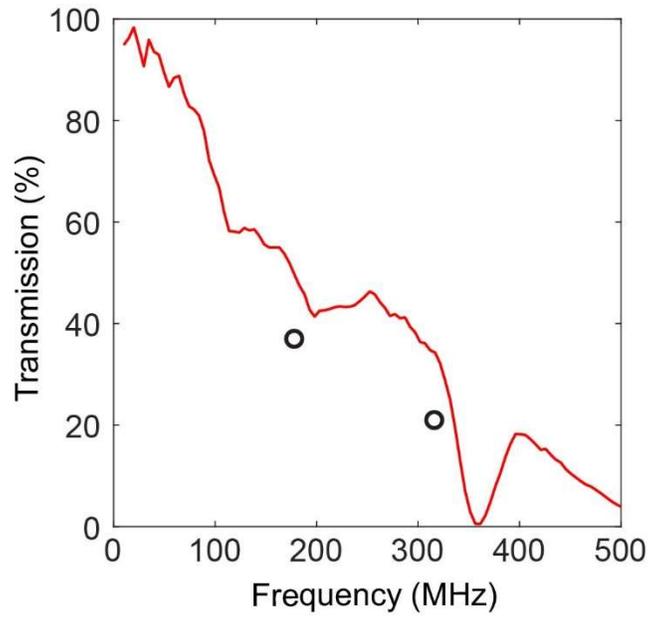

**Fig. S7: High-frequency transmission.** Red curve: VNA measurement of transmission (S12 parameter) through parts of the cabling (excluding the device) used to measure the high-frequency data in Fig. 4d. Black circles: Normalized oscillation amplitude from Fig. 4d in the main text, averaged for $\theta_x > 0$ and $\theta_x < 0$.

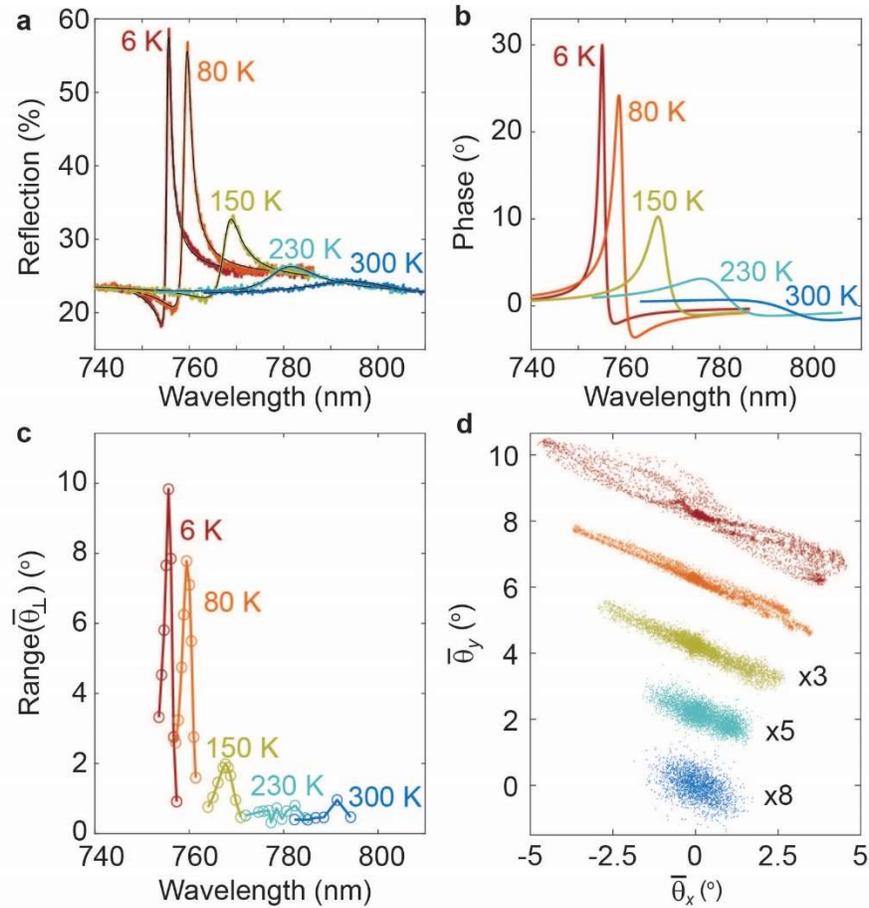

**Fig. S8: Temperature dependence. a**, Reflection spectra at temperatures 6 K (maroon), 80 K (orange), 150 K (yellow), 230 K (teal) and 300 K (blue), showing line broadening, amplitude reduction and red-shift of the exciton resonance with increasing temperature. Black: fits. **b**, Phase calculated from fits in **a**. The decrease in exciton reflection amplitude at higher temperatures reduces the available phase range of the combined exciton and background reflection. **c**, Beam deflection range perpendicular to the gate edge at the same temperatures as in **a** and **b** for a range of wavelengths near the exciton resonance. The deflection range is obtained for the same gate voltage ranges as used in the main text. The reduced phase range at higher temperatures causes a decrease in deflection range. However, the deflection range is almost unchanged from 6 K to liquid nitrogen temperature (80 K), and is still approximately 2° at 150 K. **d**, Scatter plot of beam deflection for the optimal wavelength at each temperature, displaced in the vertical direction for enhanced visibility. Plots at 150 K, 230 K and 300 K are scaled by 3, 5 and 8, respectively. Beam deflection is still observable at 230 K, and some deflection is even observed at room temperature (300 K).

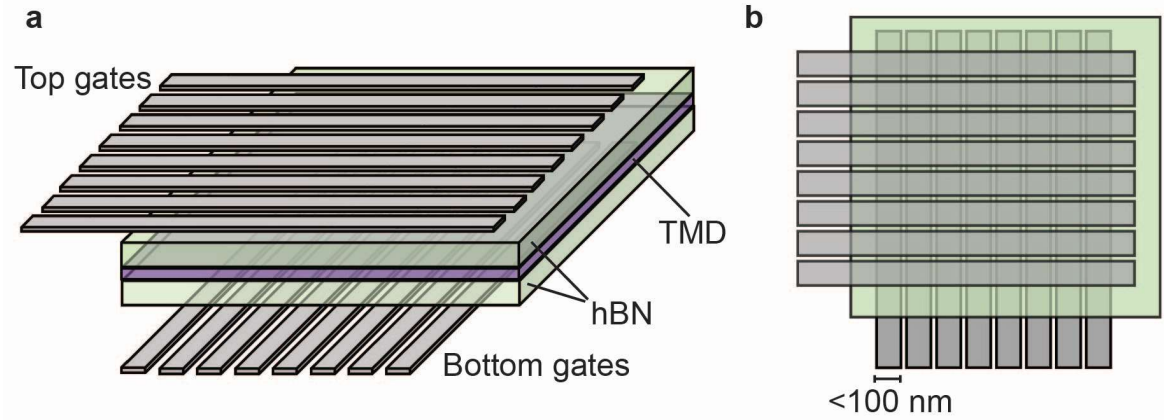

**Fig. S9: Upscaling potential. a,b,** Side and top-view schematics of one route for scaling up to devices with many pixels. Through standard etching techniques, the gates can be patterned into sub-100 nm wide strips, forming a 2D grid. By the same double capacitor mechanism as demonstrated in our work, the combination of voltages applied to top gate strip $i$ and bottom gate strip $j$ controls the phase of pixel $(i,j)$.

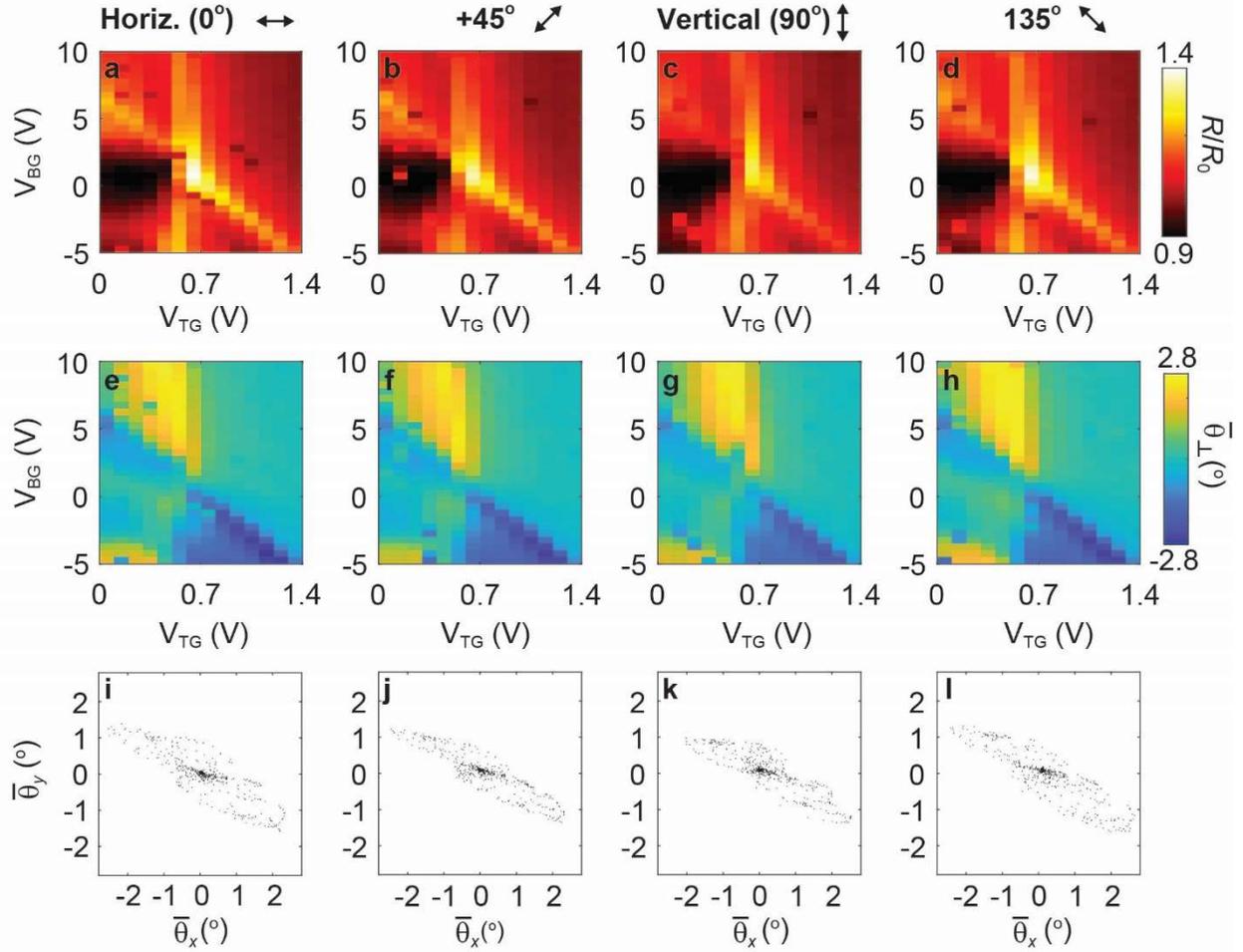

**Fig. S10: Polarization dependence of beam steering performance. a-d**, Gate dependence of total integrated reflection ($\lambda_0$=755 nm), normalized to that obtained in the highly doped regime ($V_{TG}$=1.4 V, $V_{BG}$=10 V), for polarization angles 0° (**a**), 45° (**b**), 90° (**c**) and 135° (**d**). **e-h**, Gate dependence of beam deflection perpendicular to gate edge, and (**i-l**) scatter plot of deflections for all gate combinations, for the same linear polarization angles as in **a-d**. No systematic polarization dependence of the reflection amplitude, deflection range or deflection direction is observed.

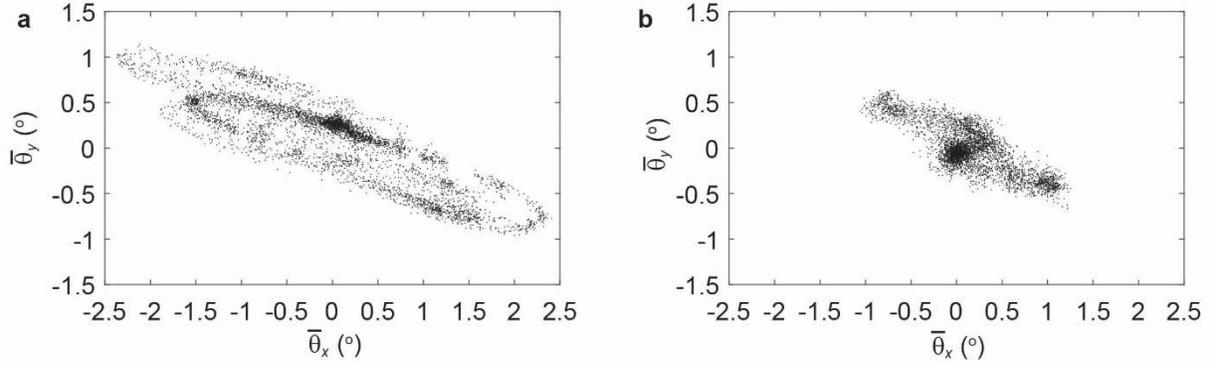

**Fig. S11: Spot size dependence. a,b**, Scatter plot of center-of-mass deflection ($\lambda_0$=754.5 nm) for the near-diffraction limited spot size used in the main text (**a**), and a ~ 2.5 times larger spot size (**b**). The deflection range is found to be approximately a factor of 2 smaller for the larger beam spot (**b**), in good agreement with the predictions in Supplementary Note IV.

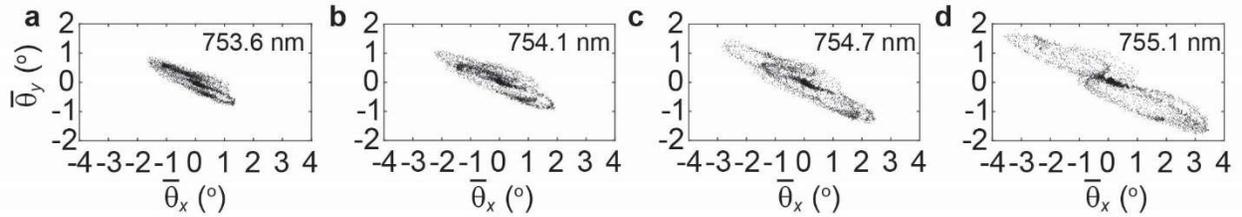

**Fig. S12: Beam steering at other wavelengths. a-d**, Scatter plot of center-of-mass deflection ($\bar{\theta}_x$, $\bar{\theta}_y$) for the same set of gate voltages as in the main text at $\lambda_0$=753.6 nm (**a**), $\lambda_0$=754.1 nm (**b**), $\lambda_0$=754.7 nm (**c**) and $\lambda_0$=755.1 nm (**d**). The deflection range is smaller at longer wavelengths, because the required large blue-shift is accompanied by a reduction in exciton reflection amplitude.

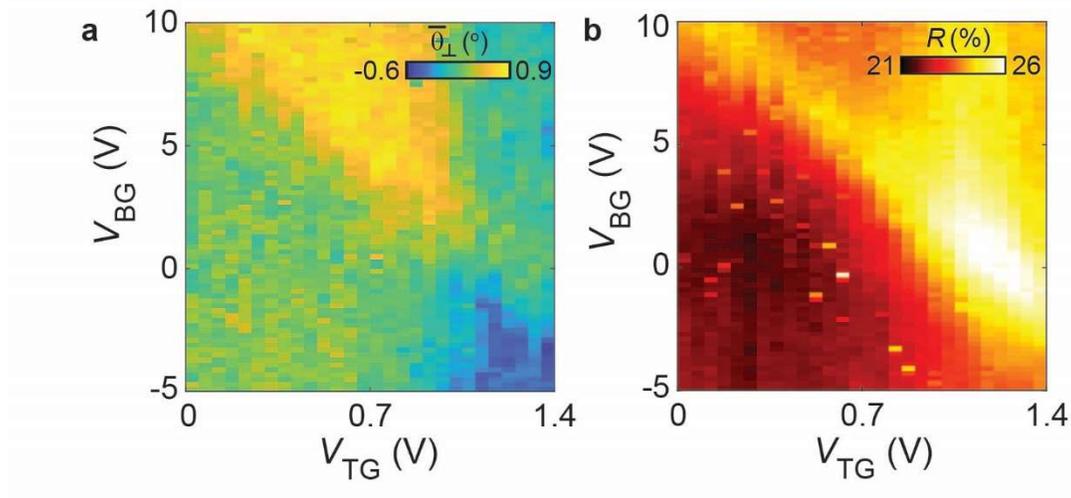

**Fig. S13: Beam steering at $\lambda_0$=752 nm. a,b,** Gate voltage dependence of deflection perpendicular to gate edge (**a**) and integrated reflection (**b**). At this short wavelength, larger voltages are required to blue-shift the exciton resonance through $\lambda_0$, compared to wavelengths closer to the intrinsic exciton resonance.

**References:**

1   Scuri, G. *et al.* Large Excitonic Reflectivity of Monolayer MoSe2 Encapsulated in Hexagonal Boron Nitride. *Physical Review Letters* **120**, 037402 (2018).
2   Zhang, B., Zerubia, J. & Olivo-Marin, J.-C. Gaussian approximations of fluorescence microscope point-spread function models. *Applied Optics* **46**, 1819-1829 (2007).